\documentclass{article}[12pt]
\usepackage{epsfig}
\usepackage{amsmath}
\usepackage{physics}
\usepackage{amsfonts}
\usepackage{amsmath,mathtools}
\usepackage{multicol}
\usepackage{amssymb}
\usepackage{wrapfig}
\usepackage{graphicx,psfrag,xspace}
\usepackage{pstricks,pst-node,pst-tree}
\usepackage{pst-all}
\usepackage{pstricks-add}
\usepackage{bbm}
\usepackage{bm}
\usepackage{float}

\def\rvec{{\bf r}}

\def\kvec{{\bf k}}
\def\he#1{$^{#1}$He}

\providecommand{\keywords}[1]
{
  \small	
  \textbf{\textit{Keywords---}} #1
}
\author{Henri Godfrin and Eckhard Krotscheck} 

\begin{document}
\title{The Dynamics of Quantum Fluids} \maketitle \author{}
\date{}
\begin{abstract}
  We review experimental and theoretical progress in the physical description
  of the dynamics of the quantum fluids \he4 and \he3. Historically,
  the elementary excitations in these systems have been identified as
  phonons and rotons and, in \he3, collective zero sound and
  spin-fluctuations.  Both recent high-precision measurements and
  theoretical methods have shown that the dynamics of these systems
  is actually very rich as will be discussed in detail in the body
  of this contribution.
\end{abstract}
\section{The Helium Liquids}

\begin{multicols}{2}
Helium ($^4$He and the less common isotope $^3$He) and hydrogen are
simple condensed matter systems, amenable to fundamental quantum
mechanical descriptions.  Helium remains liquid even at the absolute
zero of temperature, an effect explained by Heisenberg's uncertainty
principle, which has the consequence that light atoms confined in a
small volume have a large kinetic energy, as is the case in a liquid
or a solid, where each atom is confined by several surrounding atoms.
For this reason, in helium, kinetic energies are larger than the weak
interatomic potential energy. Even at the lowest temperatures,
substantial atomic motion is present, and the system remains liquid,
unless pressures on the order of 25 MPa are applied, and
solidification is achieved.  Another quantum effect plays an important
role in the behavior of liquid helium: as the temperature is reduced
below a few degrees Kelvin, the thermal de Broglie wave-length of the
atoms becomes larger than the interatomic distance.  The waves
describing the helium atoms overlap significantly, and a description
in terms of individual wave-packets of distinguishable particles
becomes inappropriate, it must be replaced by a quantum many-body
description in terms of indistinguishable particles, governed by
quantum statistics.

Quantum Mechanics and Statistical Physics
\cite{FetterWalecka,ThoulessBook} explain well the properties of
Quantum Gases, {\em i.e.\/} ensembles of non-interacting particles,
which necessarily belong to one of the two possible classes: Bosons
having an integer spin quantum number {\em s}), or fermions with
half-integer spin. $^4$He atoms ($s$=0) are bosons, and $^3$He atoms
($s$=$\frac{1}{2}$) are fermions.

The wave-functions of particles confined in a volume $V$ are quantized
plane waves characterized by their wave-vector $\kvec$. As shown by
A. Einstein, the ground state of a Bose gas is the BEC (Bose-Einstein
condensate), where all particles occupy the same microscopic state
with $\kvec$=0. The ground state of a Fermi gas is very different,
because Pauli’s principle forbids double occupancy for Fermions: the
particles form a Fermi sphere in wave-vector space, by completing
successive energy levels of increasing wave-vector, up to the Fermi
wave-vector $k_{\rm F}$. The energy $E_{\rm F}$ of the last one,
called the Fermi level, can be considerable.

The excitation spectrum $\epsilon(k)$ of these quantum gases is simply
given by the kinetic energies of the individual particles,
$\epsilon(k)=\hbar^2 k^2/2m$, where $m$ is the particle mass and $k$ the
the wave number.  For bosons, at low temperatures,
particles are excited from the common ground state. The energy
spectrum is the usual parabolic spectrum of free particles described
above. This behavior, however, is profoundly modified by interparticle
interactions, {\em i.e.\/}, as is the case in Quantum Fluids
\cite{FetterWalecka,ThoulessBook,PinesNoz,Wilks}.

The Helium quantum {\em fluids\/} are unique in the sense that they
are {\em very dense\/} and {\em very quantum\/}, making them a very
challenging problem for theory. The issue is made quantitative by
looking at the simple Lennard-Jones model of the helium fluids.  In
this model, the helium atoms interact via the interaction
\[V(r) = 4\epsilon\left[\left(\frac{\sigma}{r}\right)^{12}-
  \left(\frac{\sigma}{r}\right)^{6}\right]\equiv \epsilon v\left(\frac{r}{\sigma}\right)\]
where $\epsilon = 10.22\,$K is the well depth of the potential
and $\sigma = 2.556\,$\AA\ is the core size \cite{deBoerMichels}.

The interaction is not the very best (quantitative calculations are in
fact made using the Aziz potential \cite{AzizIII}), but it reproduces
the ground state properties of the liquids within a few percent.
Measuring all energies in units of the well depth, and all lengths in
units of the core size, $x\equiv r/\sigma$, the Hamiltonian of the
system is
\begin{eqnarray}
  &&\frac{1}{\epsilon} H({\bf x}_1\ldots,{\bf x}_N) =\\
&&-\frac{\Lambda^2}{8\pi^2}
\sum_i\nabla_{{\bf x}_i}^2
+ \sum_{i<j} v(|{\bf x}_i-{\bf x}_j|)\nonumber
\end{eqnarray}
\noindent
where 
\[\Lambda = \left(\frac{h}{ m \epsilon\sigma^2}\right)^{\frac{1}{2}}
\]
is de Boer's ``quantum parameter'' \cite{DeBoer1948} $\Lambda$, giving
the ratio between the de Broglie wavelength and a typical diameter of
the molecule.  $\Lambda\approx 2.67$ for $^4$He, $\Lambda\approx 3.09$
for $^3$He, but $\Lambda\approx 1.3-1.7$ for H$_2$, HD, D$_2$, and
$\Lambda < 0.1$ for heavy rare gases.

Moreover, the density of $^4$He at zero pressure is $\rho_0 =
0.02185\,$\AA$^{-3} = 0.365/\sigma^{-3}$ and for $^3$He $\rho_0 =
0.0163\,$\AA$^{-3} = 0.273/\sigma^{-3}$, in other words the average
particle distance is of the order of the core size.

\begin{figure}[H]
   \begin{center}
	\includegraphics[width=1.0\columnwidth,angle=-90]{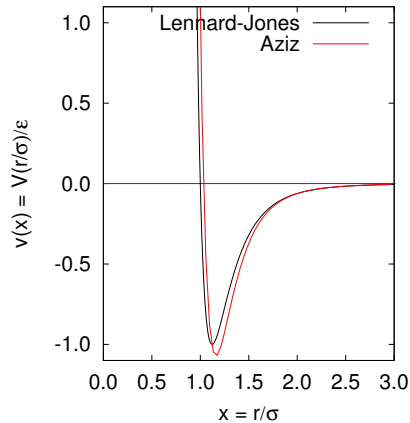}
\caption{\small{Normalized Lennard-Jones ``6-12'' interatomic potential,
    as a function of the normalized interatomic distance. The
    normalization factors for distances ($\sigma$) and potential
    energies ($\epsilon$) are given in the text. We also show, for comparison,
    the Aziz potential \cite{AzizIII} which is the currently accepted
    most accurate static 2-body interaction.
  } \label{fig:LJ}}
 \end{center}
\end{figure}

\section{$^4$He in 3D}

Liquid helium ($^4$He) is the archetype of a Bose Liquid.  Immediately
after the discovery of the superfluidity of helium
\cite{Kapitza1938,Allen1938} by Kapitza (Moscow) and Allen and Misener
(Toronto), F. London \cite{London1938} understood that the system was
undergoing Bose-Einstein condensation, and L. Tisza proposed a
``two-fluids model'' featuring a coexistence of a normal fluid and a
superfluid \cite{Wilks}.  A Bose-Einstein condensed gas is not
superfluid, however, and therefore the superfluidity observed in
helium is an additional effect, associated to the interactions.  A
major step in the understanding of the physics of helium was made by
L.D. Landau, who found a deeper interpretation of Tisza's two-fluids
model.  Landau postulated that the ``normal fraction'', which carries
the entropy, is in fact the ensemble of thermal excitations of the
superfluid ground state.  This interpretation removed the unphysical
consideration of two categories within a set of indistinguishable
atoms.  Landau considered elementary excitations of the fluid, in the
form of quantized density fluctuations (``phonons''), similar to the
Debye phonon modes of a crystalline solid.  In order to calculate the
thermodynamic properties by a Debye-like approach, where the energy is
obtained by a sum over all modes, weighted by the Bose factor, Landau
postulated that the dispersion relation $\epsilon(k)$ had a linear
part $\epsilon=\hbar ck$, where $c$ is the speed of sound, and another
part displaying a minimum, the ``roton gap'' of energy $\Delta$.
Landau's intuition was guided by specific heat measurements, showing a
cubic temperature dependence at low temperatures (below 0.5 K), as
expected from the linear part of the dispersion relation, as well a an
exponential contribution at higher temperatures (on the order of 1K),
characteristic of an energy gap.

In a first article \cite{Landauroton}, Landau proposed a spectrum with
two branches, with a parabolic branch centered at k=0 in addition to
the linear branch.  This spectrum was not found satisfactory by Landau
himself, the presence of two branches being reminiscent of Tisza's
``two kinds of atoms''.

\begin{figure}[H]
    \begin{center}
   \includegraphics[width=0.9\columnwidth]{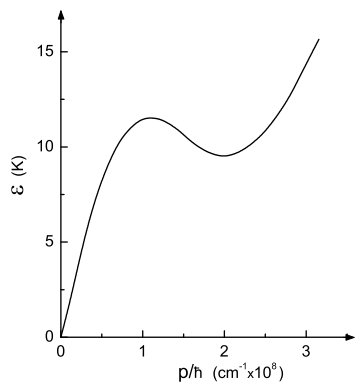}
\caption{\small{The dispersion relation proposed by L. D. Landau in
    1947 to describe the elementary excitations in superfluid $^4$He
    at zero pressure.}
		\label{fig:landau-1947} }
\end{center}
\end{figure}

In a second publication \cite{Landauroton2}, he presented his
celebrated ``Landau dispersion relation'', where a single branch of
excitations characterizes the full spectrum
(Fig.\,\ref{fig:landau-1947}).  With two parameters, the sound
velocity $c$ and the roton gap $\Delta$, both obtained from the
temperature dependence of the specific heat, Landau's theory provides
an excellent account of the experimental properties of superfluid
helium at low temperatures. The existence of a ``second sound''
collective mode, where the superfluid and normal components oscillate
with opposite phases while keeping the total density constant, as
proposed by Tisza, corresponds in Landau's model to propagating
entropy (and hence temperature) waves.

Another fundamental observation due to Landau, is that the particular
shape of the dispersion relation leads to the superfluid behavior.  A
simple argument of energy and momentum conservation shows that the
creation of excitations by the displacement of an object immersed in
the fluid is not possible, unless the object's velocity exceeds a
critical value.  The linearity of the dispersion relation at low wave
vectors is essential for this to happen: in a Bose gas, where the
dispersion relations is parabolic, the critical velocity is zero, the
system remains normal even at the absolute zero of temperature.  The
evolution of the dispersion relation from the parabolic shape
characteristic of a gas, to the linear behavior observed as the
interaction is switched on, has been explained by Bogoliubov (1947)
\cite{Bogoliubov} in the weak interaction limit.  The dispersion
relation of the elementary excitations plays therefore a major role in
the determination of the thermal properties, and also in the
superfluid nature of the liquid's ground state.

Landau's phenomenological theory introduced the fruitful concept of
``quasiparticles'', describing in a simple way the elementary
excitations of complex many-body systems.  The concept was of course
already present in the Debye theory of solids, but the generalization
to helium, a non-periodic system of strongly interacting particles,
opened a new field of physics.

There were however two serious interrogations after Landau's theory
was released.  The first was obviously related to the necessity of an
experimental observation of the proposed dispersion curve. And the
second, to the need of a microscopic theory providing some insight in
the superfluid helium behavior, and able to test the validity of
Landau's model.

Experimentally, the evidence for a Landau-like dispersion relation was
missing: only indirect evidence obtained form thermodynamic
measurement (specific heat, thermal conductivity, viscosity, fountain
pressure, etc.) was available.  Feynman and Cohen suggested the use of
inelastic neutron scattering, a technique already used to measure
phonon dispersion curves in crystals, to try to observe the excitation
dispersion curve of superfluid helium.  The first proof of the
existence of a roton minimum was reported by H. Palevsky, K. Otnes,
K.\, E. Larsson, R. Pauli and R. Stedman (1957)
\cite{Palevsky1957,Palevsky1958}, clearly showing that the elementary
excitations in superfluid helium are extremely sharp, compared to
those of normal helium.  These exciting results motivated extensive
investigations of the dynamics of helium, covering the low energy
region, {\em i.e.\/} the Landau-like dispersion relation, the
multi-excitations region at higher energies, as well as the very high
energy region of the spectrum, where BEC can be investigated
\cite{Glyde2018}.

\begin{figure}[H]
    \begin{center}
   \includegraphics[width=1.0\columnwidth]{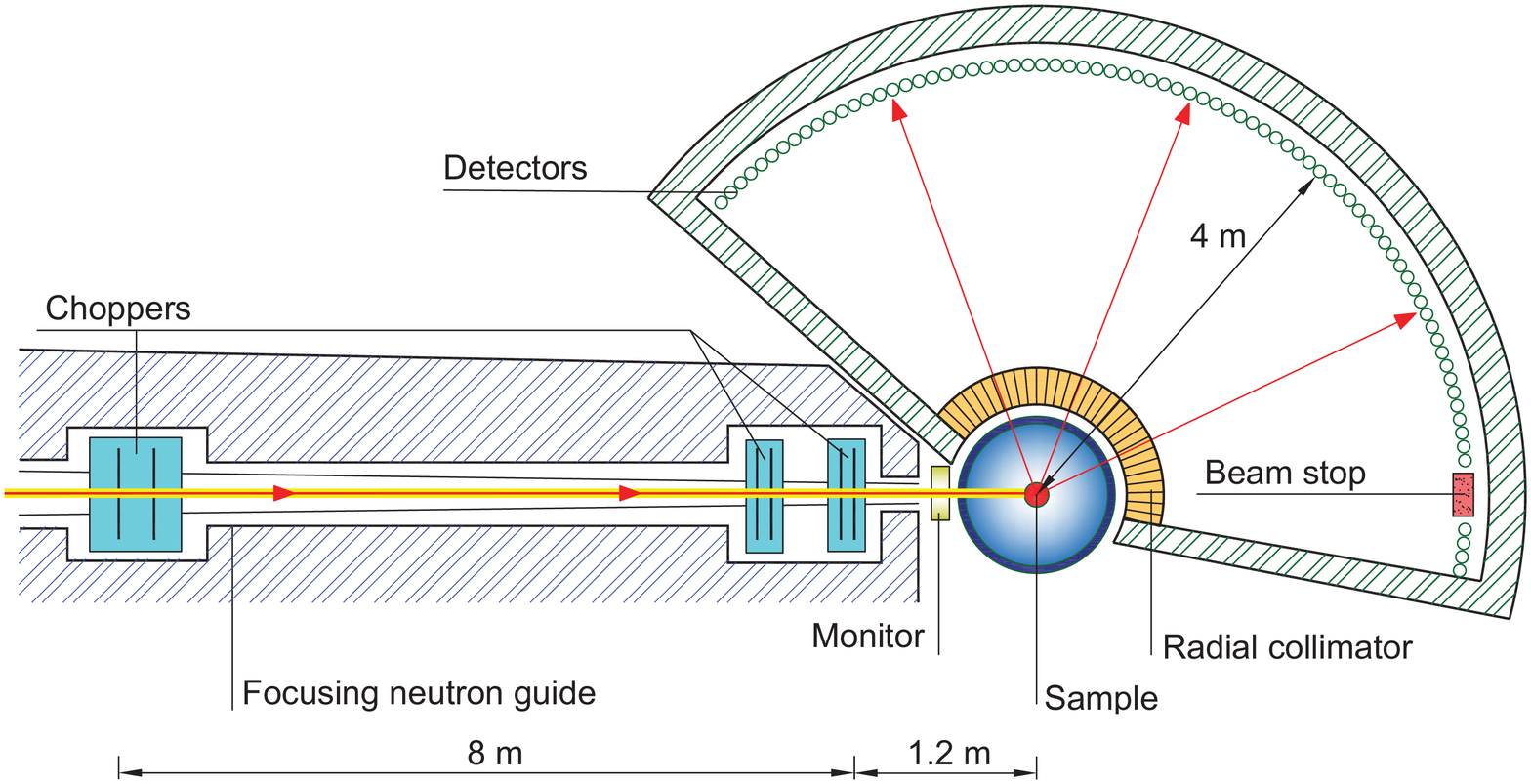}
     \includegraphics[width=1.0\columnwidth]{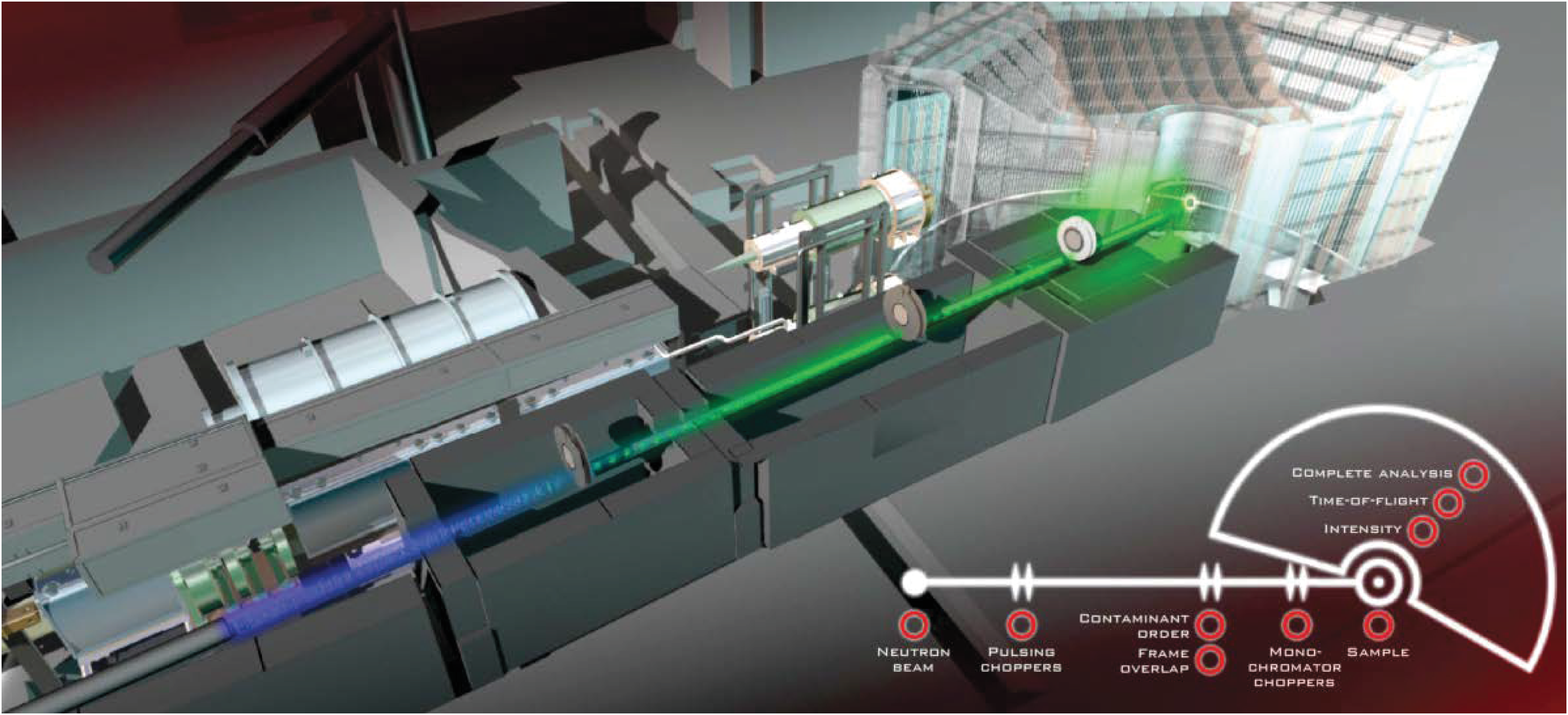}
\caption{\small{The time-of-flight neutron scattering spectrometer IN5
    at the Institut Laue-Langevin. }
        \label{fig:IN5} }
\end{center}
\end{figure}

The principle of the inelastic neutron scattering is simple, but
measurements are demanding, and high flux neutron sources (reactors,
spallation sources) are needed.  The helium sample is placed in an
incident beam of ``monochromatic'' (of a given energy) neutrons, which
are scattered by the helium sample, and then detected at some distance
from the sample, as a function of angle and time of arrival, in the
so-called time-of-flight technique.  Other methods (triple-axis,
back-scattering, spin-echo, etc.) have also been used, offering a
different coverage of the momentum-energy plane, and also different
resolutions in energy and momentum.

Energy and momentum conservation allow the determination of the energy
transfer and the momentum transfer from the neutron to the helium
sample, which correspond to the energy $\epsilon = \hbar\,\omega$ and
momentum $\hbar \vec{k}=\hbar \vec{Q}$ of the density fluctuations
created in the helium.  The incident neutron has an energy $E_i$ and a
wave-vector $\vec{k_i}$, and, after scattering, a final energy $E_f$
and a wave-vector $\vec{k_f}$; the wave-vector transfer is
$\vec{Q}=\vec{k_i}-\vec{k_f}$, and the energy transfer $\hbar\,\omega
= E_i-E_f$.
The first microscopic theory of superfluid helium, due to R. P. Feynman
(1954) \cite{Feynman}, is a wonderful example of physical
intuition. He found a general form for a variational wave-function
which successfully explained the general shape of the dispersion
relation, including the roton minimum.  It was soon completed by
Feynman and Cohen \cite{FeynmanBackflow}, who introduced, improving
the variational wave-functions, the concept of ``backflow'': the flow of
helium atoms around a moving helium atom.

Dealing with many-body correlations, including dynamic effects, has
been central to the development of microscopic theories, which have
been considerably refined in the last decades, reaching now a
quantitative power of prediction, as shown below.

\begin{figure}[H]
    \begin{center}
   \includegraphics[width=0.83\columnwidth]{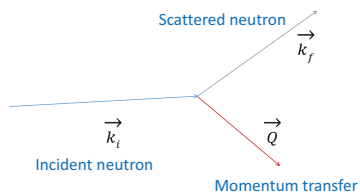}
\caption{\small{Neutron scattering: an incident neutron of wavevector $\vec{k_i}$ creates an excitation of momentum $\hbar\vec{k}=\hbar\vec{Q}$ in the superfluid. Q is the wavevector transfer (see text) } 
		\label{fig:scattering} }
\end{center}
\end{figure}

The theory of neutron scattering \cite{Lovesey,GlydeBook,Glyde2018}
relates the double differential scattering cross section per target
atom, which is the parameter measured by a scattering experiment, to
the dynamic structure factor $S(Q,\omega)$ and the dynamic
susceptibility $\chi(Q,\omega)$, the latter being the functions
calculated by microscopic theories.  They are given by the following
expression, where $b_c$ is the coherent scattering length of helium:
\begin{equation}
\label{eq:neutrons}
\frac{\partial^{2}\sigma} {\partial\Omega~\partial{E_f}} = \frac{b^{2}_{c}} {\hbar} \frac{k_f} {k_i} S(Q,\omega)  
\end{equation}
 
Results are generally presented in terms of $S(Q,\omega)$, where sharp
peaks of high intensity are observed, as expected from the creation of
Landau ``single-excitations'', as well as a more complex, broad
landscape of multi-excitations.  A vast literature on the subject
exists, as described in the review articles by Woods and Cowley,
Stirling, Glyde (see \cite{Glyde2018}), the book by Glyde
\cite{GlydeBook}, and recent works \cite{skwpress,He4Dispersion},
which benefited from the substantial progress of neutron scattering
facilities and instruments.
 
\begin{figure}[H]
\newcommand*{\vcenteredhbox}[1]{\begin{tabular}{@{}c@{}}#1\end{tabular}}
\begin{center}
  \vcenteredhbox{\includegraphics[width=0.4\textwidth]%
    {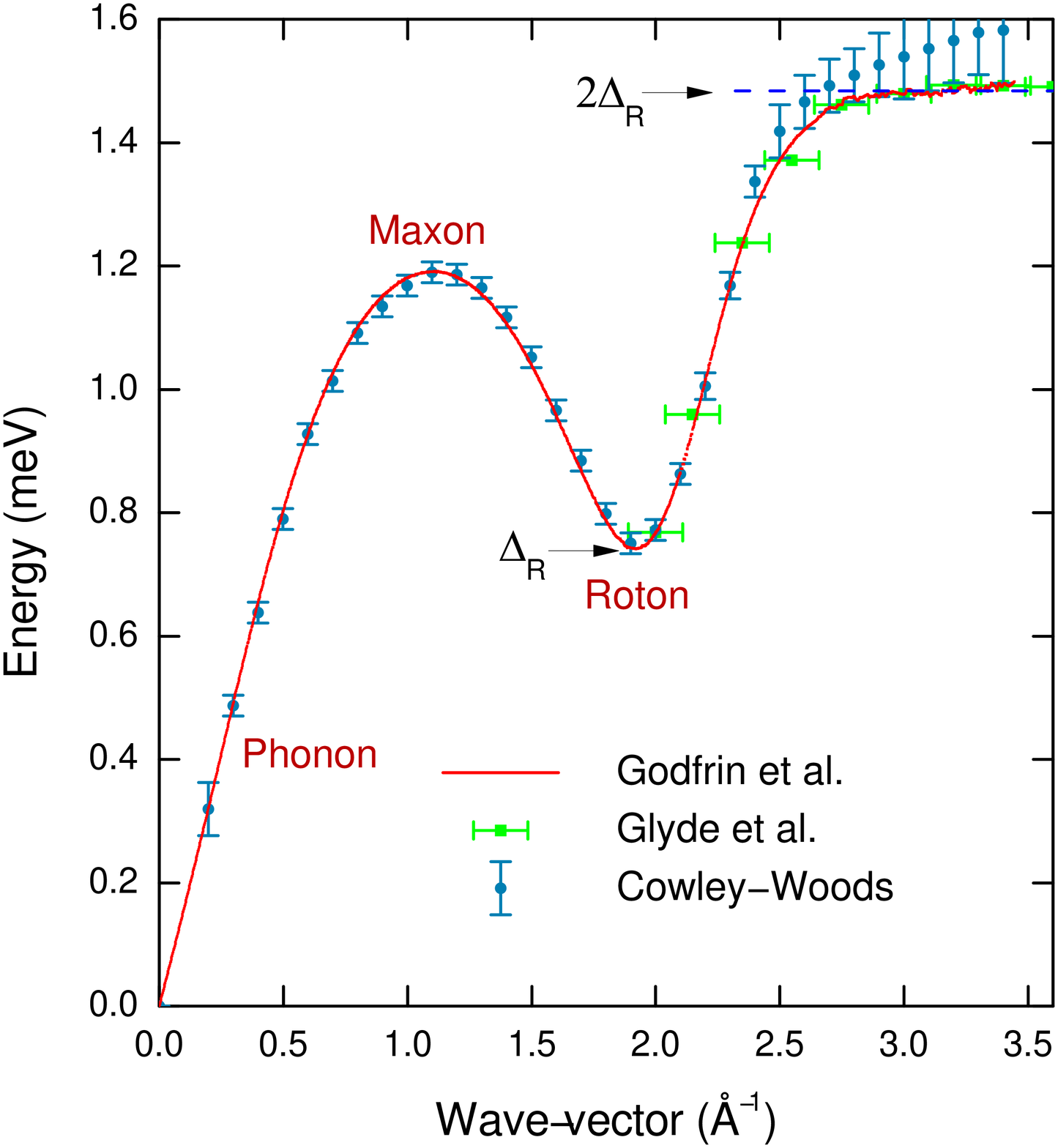}}\\
  \vcenteredhbox{\includegraphics[width=0.43\textwidth]%
    {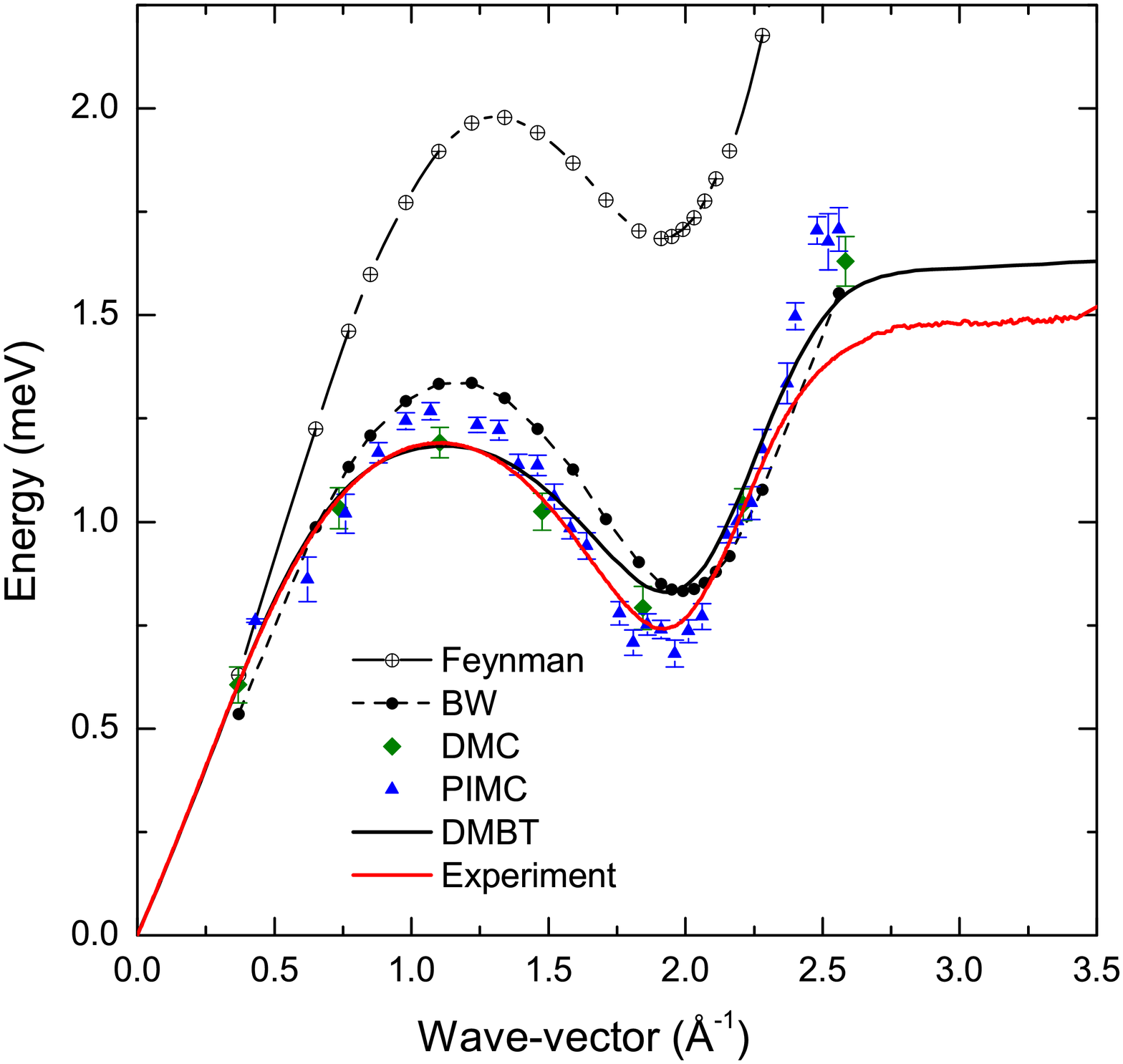}}
     \end{center}
\caption{\small{The dispersion relation $\epsilon(k)$ of excitations
    in superfluid $^4$He at zero pressure.  Top: Experimental data,
    displaying the phonon, roton, maxon, and plateau regions of the
    spectrum. The red dots are recent high-accuracy, highly pixelized
    data (seen at this scale as a continuous line, the dispersion is
    only visible at the highest wave-vectors) from
    Ref. \cite{He4Dispersion}, the blue dots with error bars are from
    the original work by Cowley and Woods \cite{CowleyWoods} and the
    green dots from the high-momentum measurements by Glyde {\em et
      al.}  \cite{Glyde-Gibbs-98}. Bottom: theoretical results, from
    Feynman's model \cite{Feynman} to modern calculations such as
    Brillouin-Wigner perturbation theory with correlated wave function
    (BW, Ref. \cite{JaFe2,LeeLee}), diffusion (DMC
    \cite{BoronatRoton,BoronatCasulleras}) and path-integral (PIMC,
    \cite{PhysRevB.93.104510}) Monte Carlo calculations and dynamic
    many-body theory (DMBT) \cite{eomIII,skwpress}, compared to the
    experimental curve \cite{He4Dispersion} \label{fig:dispersion}. }}
\end{figure}

The dispersion relation $\epsilon(k)$ of the sharp single-excitations, 
as measured in recent inelastic neutron scattering experiments, is shown in
Fig. \ref{fig:dispersion}. It closely resembles the curve predicted by 
Landau, with a linear part at low wave-vectors (``phonons''), and
a deep gap (``roton'' minimum) at a wave-vector $k$$\sim$2\AA$^{-1}$.
The name roton is historical, no special rotation takes place; the
minimum in energy reflects an incipient localization (short-range
order), which eventually will lead to a liquid-solid transition under
a pressure $P$$\sim$2.5\,MPa.  The dispersion curve becomes flat for
$k$$\geq$2.8\,\AA$^{-1}$ as the energy reaches twice the roton gap,
forming ``Pitaevskii's plateau''.

The calculation of the low temperature ($T<1.3\,K$) thermodynamic
properties of helium at low temperatures only involves the statistical
physics of phonons and rotons of low energy. That is, the properties
of the strongly interacting bosonic system are simply given by those
of a collection of non-interacting bosonic ``quasi-particles''
described by the phonon-roton curve \cite{He4Dispersion}. At high
temperatures, where the number of rotons becomes very large,
roton-roton interactions start playing a role. At the ``lambda point''
(T=2.17\,K), $^4$He becomes a normal fluid.

Theoretical studies of the dynamic structure function in \he4 began
with the work of Feynman \cite{Feynman} and Feynman and Cohen
\cite{FeynmanBackflow}, the Feynman theory of elementary excitations
was developed in a systematic Brillouin-Wigner perturbation theory by
Jackson and Feenberg \cite{JaFe,JaFe2,JAC69,JacksonSkw} in terms of a
basis of Feynman excitation states
\begin{equation}
  \left\{\left|\kvec\right\rangle\right\}
  = \left\{S^{-1}(k)\hat\rho_{\kvec}\left|\Psi_0\right\rangle\right\}
\label{eq:FeynmanStates}
\end{equation}
where $\hat \rho_\kvec$ is the density operator and
$\left|\Psi_0\right\rangle$ is the ground state.  An important
contribution was the identification of classes of theories for the
dynamic structure function \cite{JacksonSumrules} that satisfy the
$\omega^0$ and $\omega^1$ sum rules exactly. The most complete
evaluation of the phonon–roton dispersion relation in terms of
Brillouin–Wigner perturbation theory was carried out by Lee and Lee
\cite{LeeLee} who obtained an impressive agreement with the
experimental phonon–roton spectrum up the wave number of $Q =
2.5${\AA}$^{-1}$. The major drawback with these old calculations was
that the required input -- pair- and three-body distribution functions
-- were poorly known.  Manousakis and Pandharipande \cite{IndianSkw}
tried to generalize the input states of the Brillouin-Wigner
perturbation theory to include ``backflow'' correlations in the spirit
of Feynman and Cohen \cite{FeynmanBackflow}. Through the gradient
operator acting on the wave function, in principle dynamic
correlations are introduced to all orders. The ``backflow–function''
is, however, chosen per physical intuition rather than by fundamental
principles, and the evaluation of the perturbative series becomes very
complicated. Topologically, diagrams similar to those of Lee and Lee
\cite{LeeLee} were included.  While the accuracy of the theoretical
roton energy is comparable to that of Lee and Lee, one can clearly see
an inconsistency since the energy of the Pitaevskii-plateau
\cite{Pitaevskii2Roton} lies below twice the energy of the roton gap.
More recent progress \cite{eomI,eomIII} used a hybrid approach of
Brillouin–Wigner perturbation theory and equations of motion for
time–dependent multiparticle correlation functions to derive a self-
consistent theory of the dynamic density–density response of \he4.

In a  nutshell, the Feynman theory of excitations is generalized
by writing the dynamic wave function as
\end{multicols}
\begin{subequations}
  \begin{eqnarray}
  \Psi(\rvec_1,\ldots,\rvec_N;t)
  &=&
  \exp(U(t))\Psi_0(\rvec_1,\ldots,\rvec_N)\\
  U(t) &=&
  \sum_i \delta u_1(\rvec_i;t) +  \sum_{i<j} \delta u_2(\rvec_i,\rvec_j;t)\
  +\ldots\,.
  \label{eq:JacksonFeenbergWaveFunction}
\end{eqnarray}
\end{subequations}
\begin{multicols}{2}
The Feynman form of the wave function is obtained by omitting all
$n$-body fluctuations $\delta u_n(\rvec_1,\ldots\rvec_n;t)$ for $n\ge
2$. The different theoretical descrptions basically differ by the way
the $n$-body fluctuations are determined.
  
Recent novel numerical methods
\cite{BoronatCasulleras,BoronatRoton,PhysRevB.82.174510,PhysRevB.88.094302,NGMV2013}
give access to dynamic properties of quantum fluids. These are
important algorithmic developments that may ultimately aid in the
demanding elimination of background and multiple-scattering events
from the raw data. Of course, it is generally agreed upon that the
model of static pair potentials like the Aziz interaction
\cite{AzizIII} describes the helium liquids accurately. Hence, given
sufficiently elaborate algorithms and sufficient computing power, such
calculations must reproduce the experimental data. The aim of the
works cited above is different: The identification of physical effects
like phonon-phonon, phonon-roton, roton-roton, maxon-roton \ldots
couplings that lead to observable features in the dynamic structure
function is, from simulation data, only possible a-posteriori whereas
the semi-analytic methods permit a direct identification of these
effects, their physical mechanisms, their relationship to the ground
state structure, and their consequences on the analytic properties
of the dynamic structure function, directly from the theory.

\begin{figure}[H]
   \begin{center}
	\includegraphics[width=1.0\textwidth]{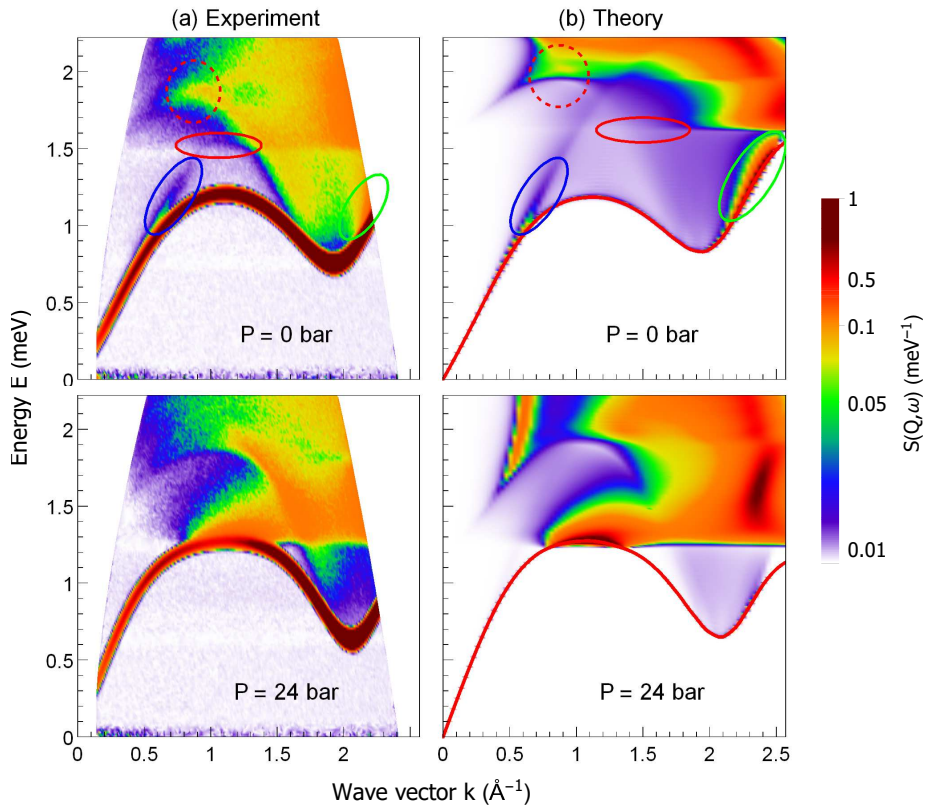}
\caption{\small{Experimental (left) and theoretical (right)
    determinations of the dynamic structure factor of superfluid
    $^4$He at zero pressure (upper graphs) and near-solidification
    pressure (lower graphs) \cite{eomIII,skwpress}. In addition to the the intense
    single-excitations dispersion seen in
    Fig.\,\ref{fig:dispersion} (red lower curves), much lower intensity
    multi-excitations (see color scale) can be observed here. Colored
    ellipses indicate excitation couplings described in the
    text.} \label{fig:4HeTheoExp}}
 \end{center}
\end{figure}

The dynamic structure of $^4$He is actually quite rich as seen in
Fig.\,\ref{fig:4HeTheoExp} \cite{eomIII,skwpress}:
``multi-excitations'' are observed at energies above the
single-excitation dispersion curve discussed above.  They can be
related to combinations of single-excitations, some of them are
indicated in Fig. \ref{fig:4HeTheoExp} by colored ellipses: in blue,
an extension of the linear phonon region (``ghost phonon'') deep into
the continuum; in red, an extension of the Pitaevskii-Plateau to small
wave-vectors; in green, above the roton minimum, a broadening of the
dispersion relation due to a Cherenkov effect; a red dashed ellipse at
high energies indicates a region of maxon-roton
couplings. Multi-excitations are well reproduced by recent
theories. The mechanisms behind these are mostly the same: It is
kinematically possible that a perturbation of energy/momentum
$(E,\kvec)$ can decay into two sharp quasiparticle excitations. For
example, the ``ghost phonon'' is caused by the effect that a
perturbation can decay into two almost collinear phonons, hence the
effect disappears at about twice the energy and wave number at which
the dispersion relation $\epsilon(k)$ ceases to be linear. Similarly,
the ``Pitaevskii-Plateau'' is caused by the effect that a perturbation
of energy $2\Delta$ can decay into two rotons of energy $\Delta$ and
wave number $k_\Delta$. Hence, energy and momentum conservation
dictate that the plateau ends at a wave number of $2 k_\Delta$, in
this case the two roton wave vectors are parallel.  Perturbations with
the same energy and smaller wave numbers can decay into two rotons
that are not collinear.  That means that the plateau can in principle
extend to zero wave number; in that case the perturbation decays into
two anti-parallel rotons.

The Cherenkov effect is particularly interesting; it is caused by the
fact that the group velocity of a quasiparticle excitation beyond the
roton ($R_+$ roton) is larger than the sound velocity. The effect is
observed only at low pressures.

\end{multicols}

\section{$^3$He in 3D}

\begin{multicols}{2}

The investigation of the dynamics of the $^3$He Fermi Liquid has
naturally followed that of the $^4$He Bose fluid
\cite{Dobbs,GlydeBook}.  From the theoretical point of view, the
problem is notoriously difficult, due to the antisymmetry requirement
for the wave-functions of many-body fermionic systems
\cite{FetterWalecka,ThoulessBook,PinesNoz}.  Obtaining accurate
results for the ground state is a challenge, even using very
sophisticated variational wave functions.

The dynamics is treated
analogously to the boson case \eqref{eq:JacksonFeenbergWaveFunction};
the local excitation functions $\delta u_n(\rvec_1,\ldots,\rvec_n;t)$
are replaced by particle-hole operators
\begin{eqnarray}\sum_{\genfrac{}{}{0pt}{1} {p_1,\ldots, p_n}{h_1\ldots h_n}}
  u_{p_1,\ldots, p_n;h_1\ldots h_n}(t)\times\nonumber\\
  {\vspace{-1cm}\quad\times a^\dagger_{p_1}\dots
  a^\dagger_{p_n}a_{h_n}\dots a^\dagger_{h_1}}\nonumber\end{eqnarray}
  where the $p_i$ are
  the quantum numbers of unoccupied (``particle'') states and the
  $h_i$ those of occupied (``hole'') states. Restricting the
  excitation amplitudes to one-particle-one-hole components leads to a
  correlated version of the random phase approximation (RPA) \cite{rings},
  the quantitative understanding of the experiments requires at least
  2-particle-2-hole amplitudes, see Refs. \cite{2p2h,eomIV}.

From the experimental point
of view, even though the same inelastic neutron techniques
successfully applied to superfluid $^4$He can be used to investigate
liquid $^3$He, in practice the experiments are extremely difficult due
to the huge neutron absorption cross-section of the $^3$He nucleus
\cite{GlydeBook}.  In typical $^4$He experiments, sample thicknesses
are larger than 1\,cm, while absorption limits the maximum thickness
in a liquid $^3$He measurement, to about 0.1\,mm. In addition, due to
the lower excitation energies, experiments in $^3$He must be performed
at much lower temperatures ($\approx 0.1$\,K) than in $^4$He ($\approx
1$\,K), to remain in the low temperature limit, where few
excitations are present. For these reasons, in spite of the
considerable interest in strongly interacting fermions, accurate
theoretical and experimental results have been available only recently
\cite{GFvDG00}.

\begin{figure}[H]
  \begin{center}
	\includegraphics[width=0.95\columnwidth,angle=0]{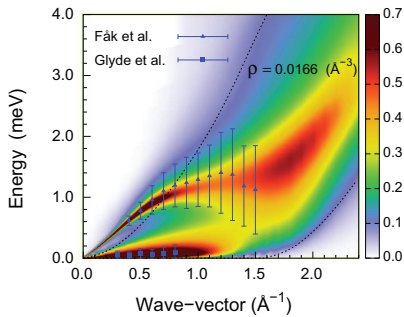}
\caption{\small{(Color online) Dynamic structure factor of bulk liquid
    $^3$He. Measured density excitations \cite{Fak} (triangles, upper
    branch) and spin-density excitations \cite{GFvDG00} (squares,
    lower branch) are clearly visible (bars on the data points
    indicate the observed width of the branch). Color intensity map,
    from light blue to dark red: DMBT microscopic theory including exchange
    \cite{eomIV}. The solid lines indicate the boundaries of the
    particle-hole band.  } \label{fig:3D-3He}}
 \end{center}
\end{figure}

In order to understand the dynamics of a Fermi {\em liquid\/}, we
begin by considering the excitations of a Fermi {\em gas\/}. In the
latter, excitations are created by removing a particle from an
occupied state inside the Fermi sphere, and placing it outside the
sphere, in a higher energy free state. The resulting states are
confined to a region of the energy {\em vs\/} wave-vector space called
the ``particle-hole band''.  The interacting fermionic system, {\em
  i.e.\/} the Fermi Liquid, behaves in a similar way.  A particle-hole
band can be observed in bulk $^3$He (see Fig \ref{fig:3D-3He}), but in
addition, collective modes are present. A density mode, named
``zero-sound'' or ``collision-less sound'', is observed.  Its physical
origin, as noted by D. Pines \cite{PinesPhysToday}, resides in the
strong interactions, and hence this mode is analogous to the
phonon-roton mode described above for bosons; statistics play here a
minor role. Its substantial energy width results from the possibility
to decay into particle-hole excitations as it enters the particle-hole
band (Landau-damping mechanism).

The thermodynamic properties of bulk liquid $^3$He are often
described, at very low temperatures, by the phenomenological Landau
Fermi Liquid model, where the properties of the interacting system are
related to those of the Fermi gas by the introduction of an
interaction function (``quasiparticle interaction''), leading to a
renormalization of parameters (effective mass, susceptibility
enhancement, etc.).  Landau theory predicts, among other physical
effects, the existence of collective modes, the zero-sound and
paramagnon modes.

Landau's Fermi Liquid Theory is only applicable at very low temperatures.
In liquid $^3$He, where the Fermi energy, expressed in temperature units, 
is of the order of several kelvins, the theory is valid below 0.1\,K. 

The reason becomes clear if one observes the actual excitation
spectrum shown in Fig.\,\ref{fig:3D-3He}.  The low energy mode,
observed below the zero-sound excitation, is a spin-density mode,
completely immersed in the particle-hole band, and hence severely
Landau-damped.  The origin of this mode, analogous to a dampened
spin-wave and often called ``paramagnon'', can be traced back directly
to the antisymmetry of the many-body wave function.

There are also some  interesting X-fay studies of the dynamic
structure function at high momentum transfers \cite{Albergamo} that
have led to a discussion about the location of the particle-hole band
\cite{Albergamo_comment,Albergamo_reply}. The authors of
Ref. \cite{Albergamo} state that {\it The obtained results show
  no evidence of such a decay: the zero-sound mode remains well
  defined in the whole explored wave number range.\/}  According to that,
the particle-hole continuum should be much lower than the continuum
of the non-interacting Fermi system. An analysis of both the data
and the theoretical predictions within DMBT \cite{Xray} shows that
there is, when the theoretical data are convoluted with the
experimental resolution, actually no contradiction to the
conventional interpretation of the scenario.
\begin{figure}[H]
 \centerline{\includegraphics[width=0.8\columnwidth]{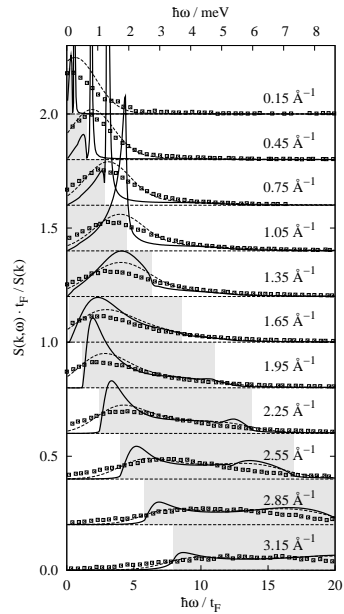}}
\caption{The figure shows the normalized dynamic structure function
  $S(k,\omega)/S(k)$ of \he3 at a density of $\rho=0.0166 $\AA$^{-3}$,
  for a sequence of wave numbers. The theoretical results obtained by
  DMBT \cite{Xray} are compared with the X-ray scattering data
  \cite{Albergamo} (squares) at saturated vapor pressure. Also shown
  are the theoretical results convoluted with the experimental
  resolution (dashed lines). The gray-shaded area shows the
  particle-hole continuum of a non-interacting Fermi fluid. $t_{\rm F}
  = \hbar^2 k_{\rm F}^2/2m$ is the Fermi energy of the non-interacting
  system.\label{fig:stack} }
\end{figure}

Details are shown in Fig.~\ref{fig:stack}.  To facilitate the
comparison with experiments, the theoretical spectra were convoluted
with the experimental resolution.  Also, the results were scaled by
$1/S(k)$ such that the integrated strength is 1 for all momentum
transfers.  The same scaling was applied to the experimental data.

\section{$^4$He in reduced dimensions}

The dynamics of quantum fluids has also been investigated in reduced
dimensions. Two-dimensional (2D) systems are obtained experimentally
by adsorption of gases onto solid substrates, usually
graphite. One-dimensional (1D) systems can also be created, they are
obtained by confining the gases inside silica or graphite nano-tubes
\cite{WadaJLTP1d,Kato87,1Dhe4}. There is also theoretical interest
\cite{StanCole,StanColeJLTP1d,PhysRevLett.116.135302,KroMi99,GBC00} in
the properties of $^{4}${\rm He} in {\it one}-dimension.

Extensive theoretical studies have clarified the nature of the
excitations in such systems. Superfluid $^4$He films of atomic
thicknesses, for instance, display in-plane density modes, as well as
capillary waves.

Let us first discuss the dynamics of rigorously 2D \he4. Basically we see
the same effects as in 3D: a linear ``phonon'' branch as well as, at high
momentum transfers, a ``roton minimum''. 

\end{multicols}
\begin{figure}[H]
\centerline
     {\includegraphics[width=0.33\columnwidth,angle=-90]{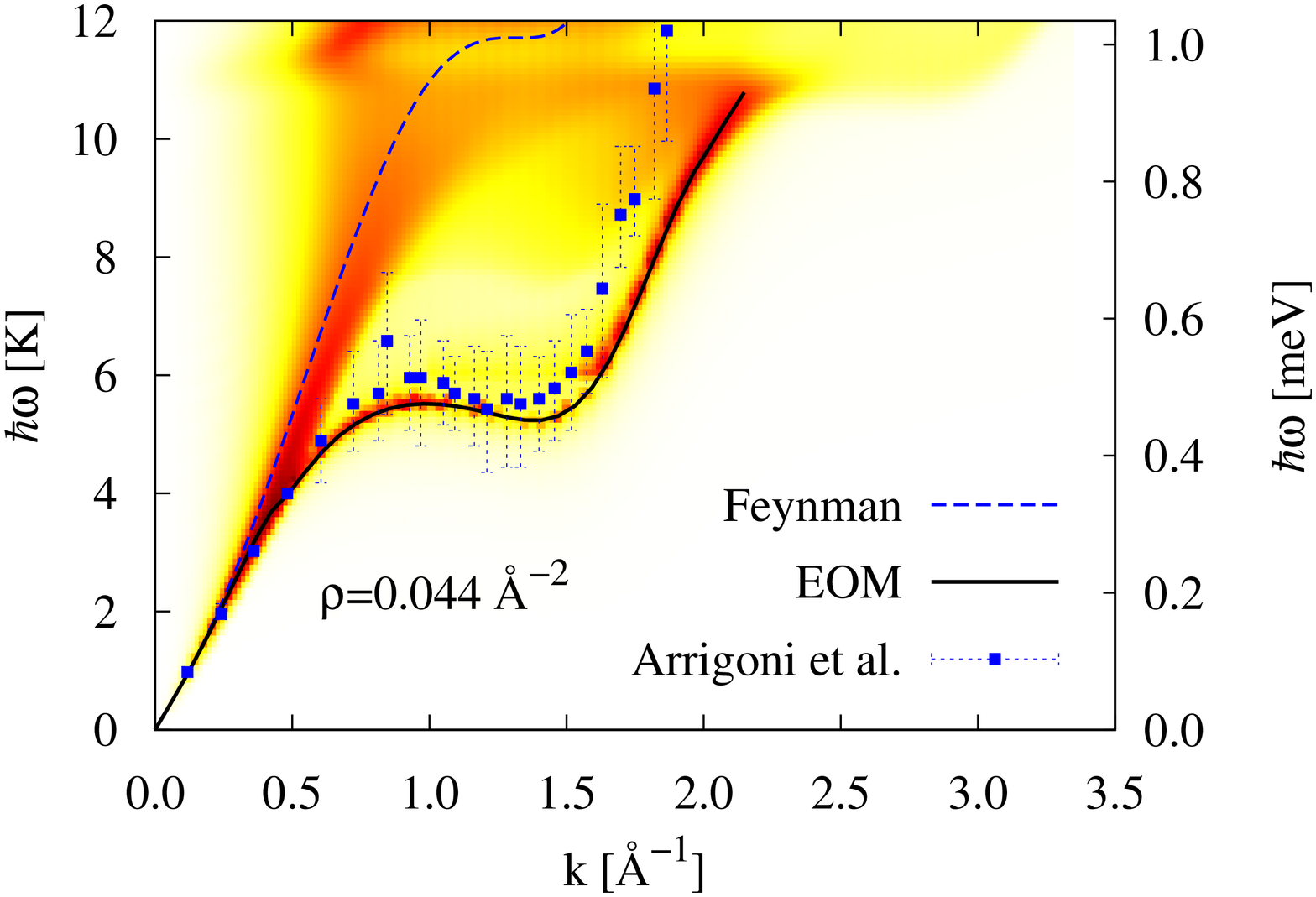}
     \includegraphics[width=0.33\columnwidth,angle=-90]{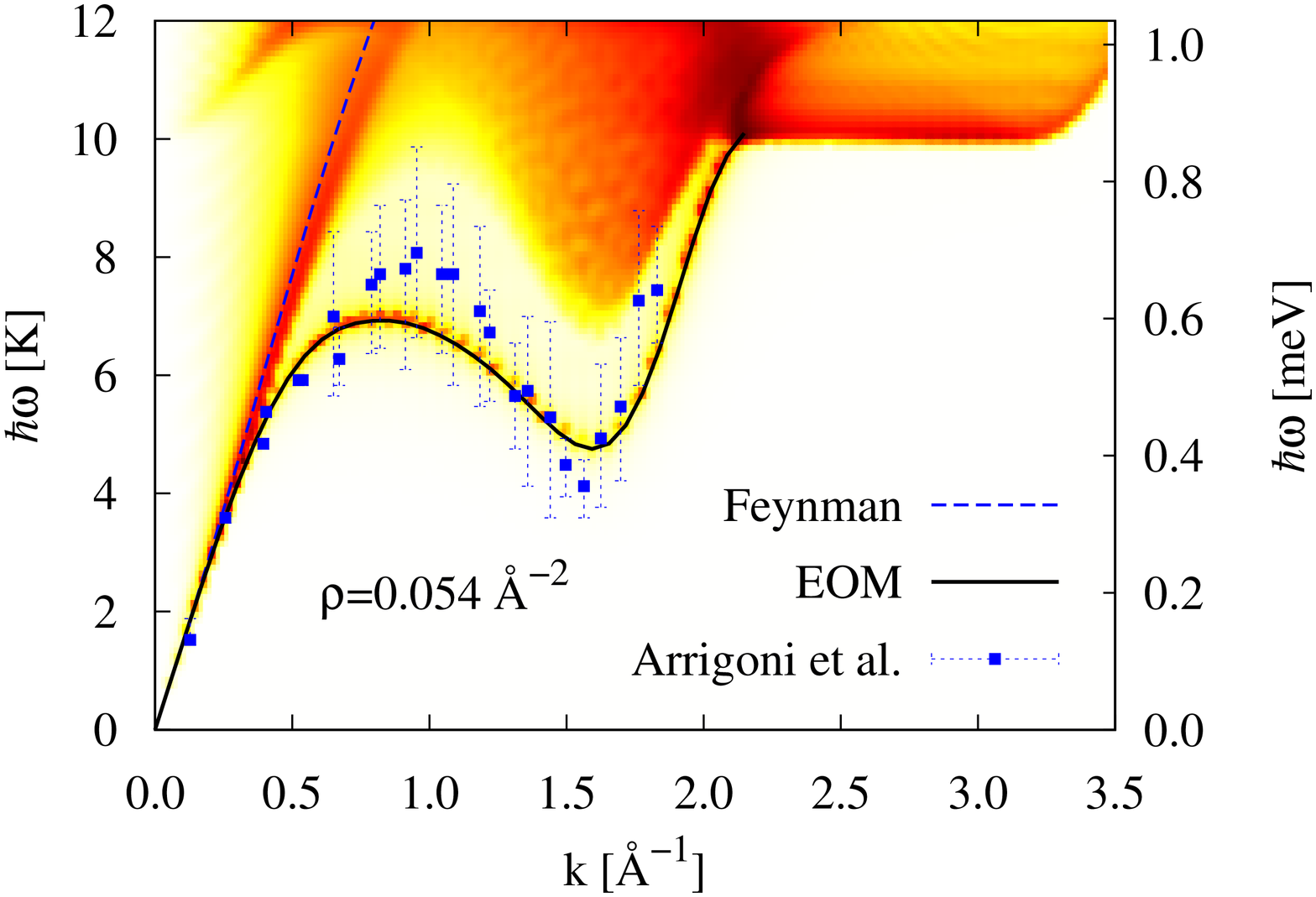}}
\centerline
    {\includegraphics[width=0.33\columnwidth,angle=-90]{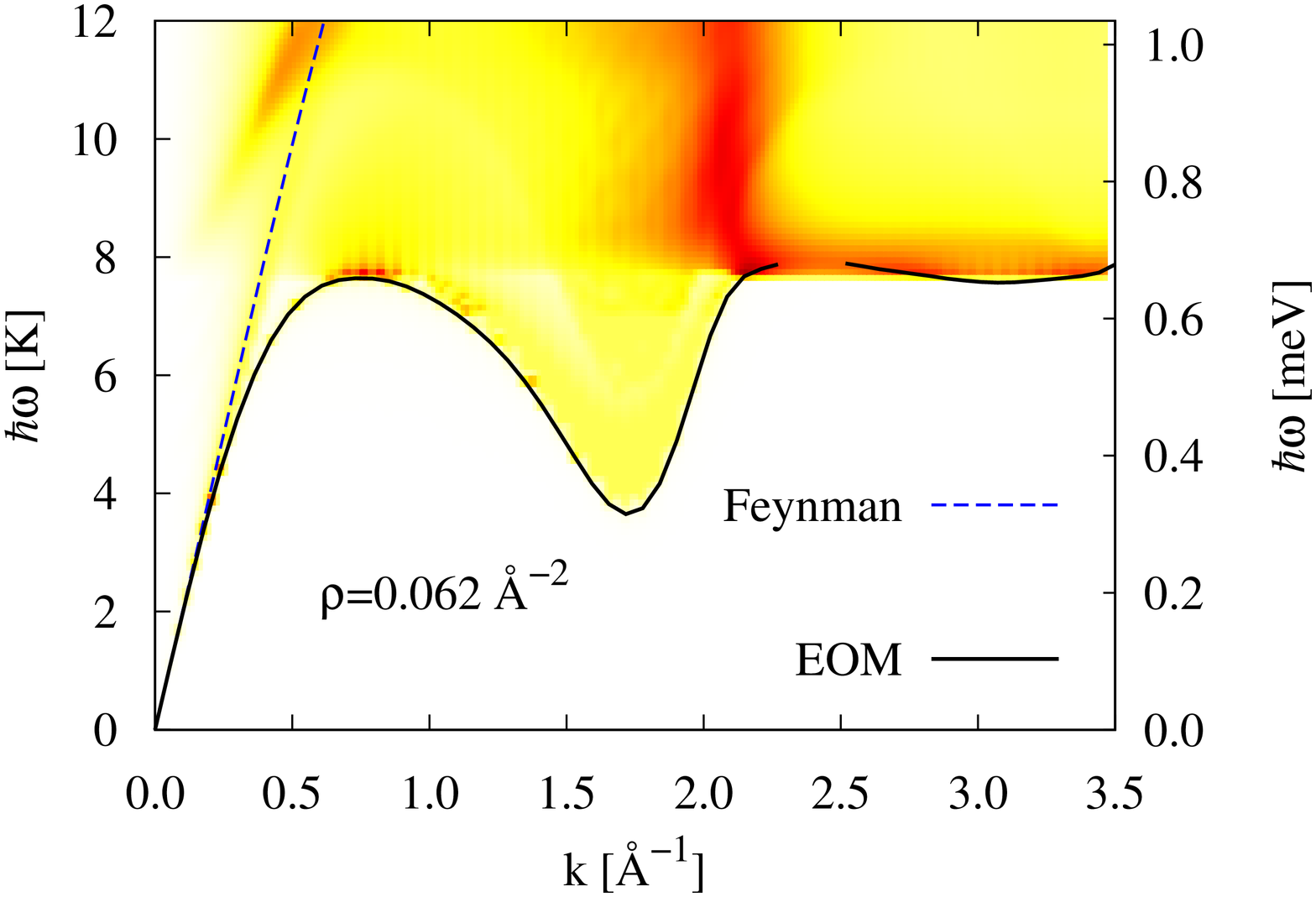}
      \includegraphics[width=0.33\columnwidth,angle=-90]{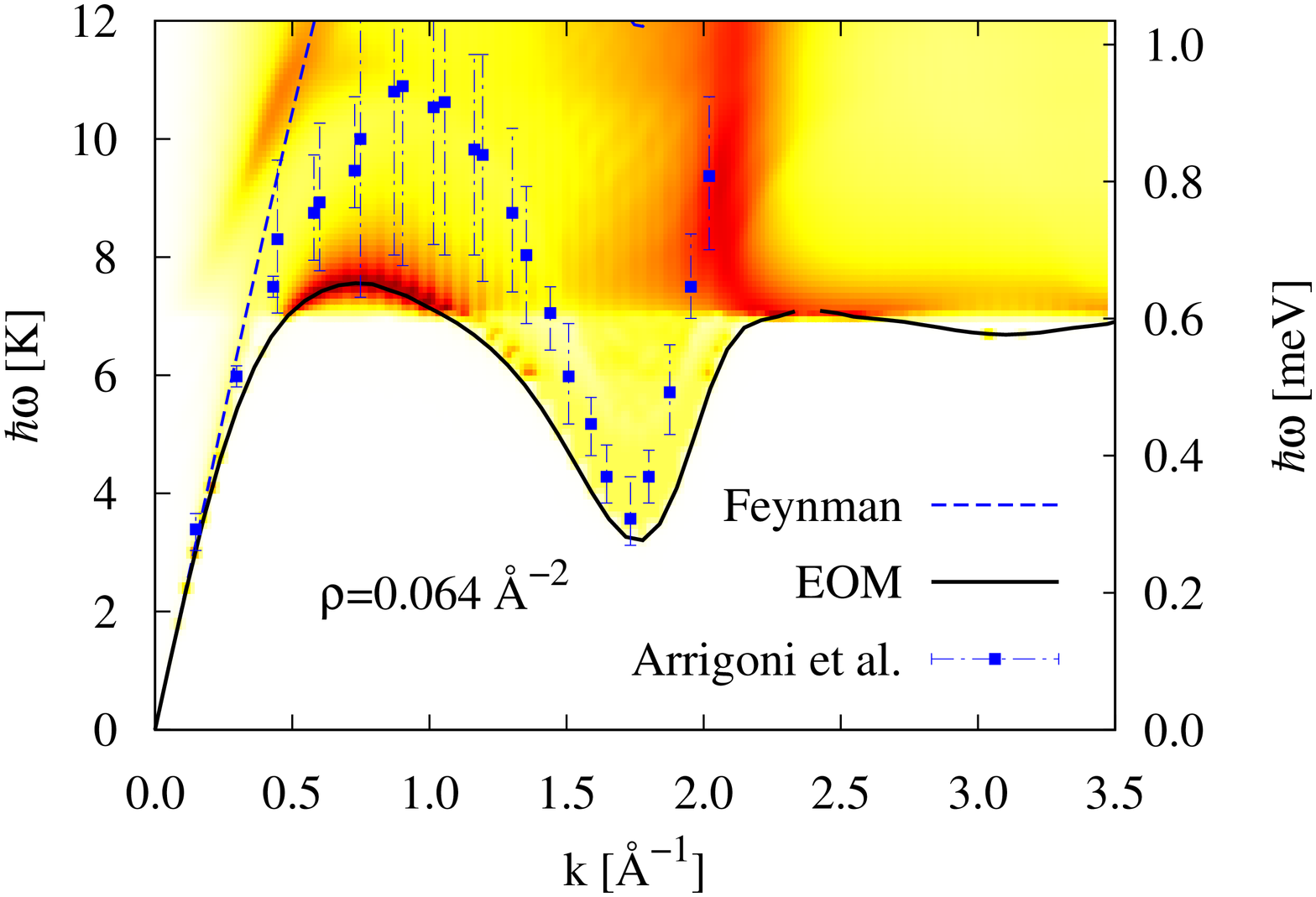}}
    \caption{(Color online) The figure shows contour plots of the
      dynamic structure function of two-dimensional liquid \he4
      for a sequence of areal densities as shown
      in the legends. The colors have been chosen to highlight the
      prominent features, darker colors correspond to higher values of
      $S(k,\hbar\omega)$. 
      We also show the Feynman spectrum, and the simulation
      data of Ref. \cite{Arrigoni2013}. From Ref. \cite{2dhe4}.}
\label{fig:skwplots}
\end{figure}
\begin{multicols}{2}
One can also see the somewhat finer details discussed above: At low
densities, the ``ghost-phonon'', and at high densities the plateau
coming down and extending to long wave lengths. A second striking
feature is the appearance of a sharp mode below the plateau.  We
stress the difference: normally, the plateau is a threshold above
which a wave of energy/momentum $(\hbar\omega, k)$ can decay into two
rotons. This has the consequence that the imaginary part of the
self-energy $\Sigma(k,\hbar\omega)$ is a step function and the real
part has a logarithmic singularity \cite{Pitaevskii2Roton}. A
collective mode is, on the other hand, characterized by a singularity
of the $S(k,\hbar\omega)$.  Figs. \ref{fig:skwplots} show, for the two
highest densities, the appearance of a sharp discrete mode {\em
  below\/} the plateau. At a wave number of $k\approx
2.6\,$\AA$^{-1}$, the collective mode is still merged into the
continuum. With increasing wave number, we see, however, a clearly
distinguishable but rather weak ($Z(k)\approx 0.02$) mode about 0.3 K
below the plateau. We have mentioned already above that the roton
should be seen as an emergence of short--ranged order, the ``ghost of
a Bragg Spot'' \cite{NozSolid,nozieres2006}. If this is the case, then
one might see a second Bragg spot. Such a thing is not seen, but
location of the secondary minimum in 2D corresponds indeed to the
position of a second Bragg spot of a triangular lattice.

Density modes are analogous to the phonons and rotons already
mentioned in the case of bulk helium, but due to the adsorption
potential of the substrate, the fluid is inhomogeneous, and
solidification of the first two or three atomic layers is observed in
thick films. Layered-excitations are observed in this regime.

Capillary waves, also called ``ripplons'', have a different nature:
these are surface waves, which do not (in first approximation)
correspond to compression/expansion of the fluid, but to a change of
height in the external potential (gravitational in the case of a bulk
helium surface, or substrate potential in adsorbed films).

\begin{figure}[H]
  \vspace{-0.5cm}
   \begin{center}
	\includegraphics[width=0.42\textwidth,angle=-90]{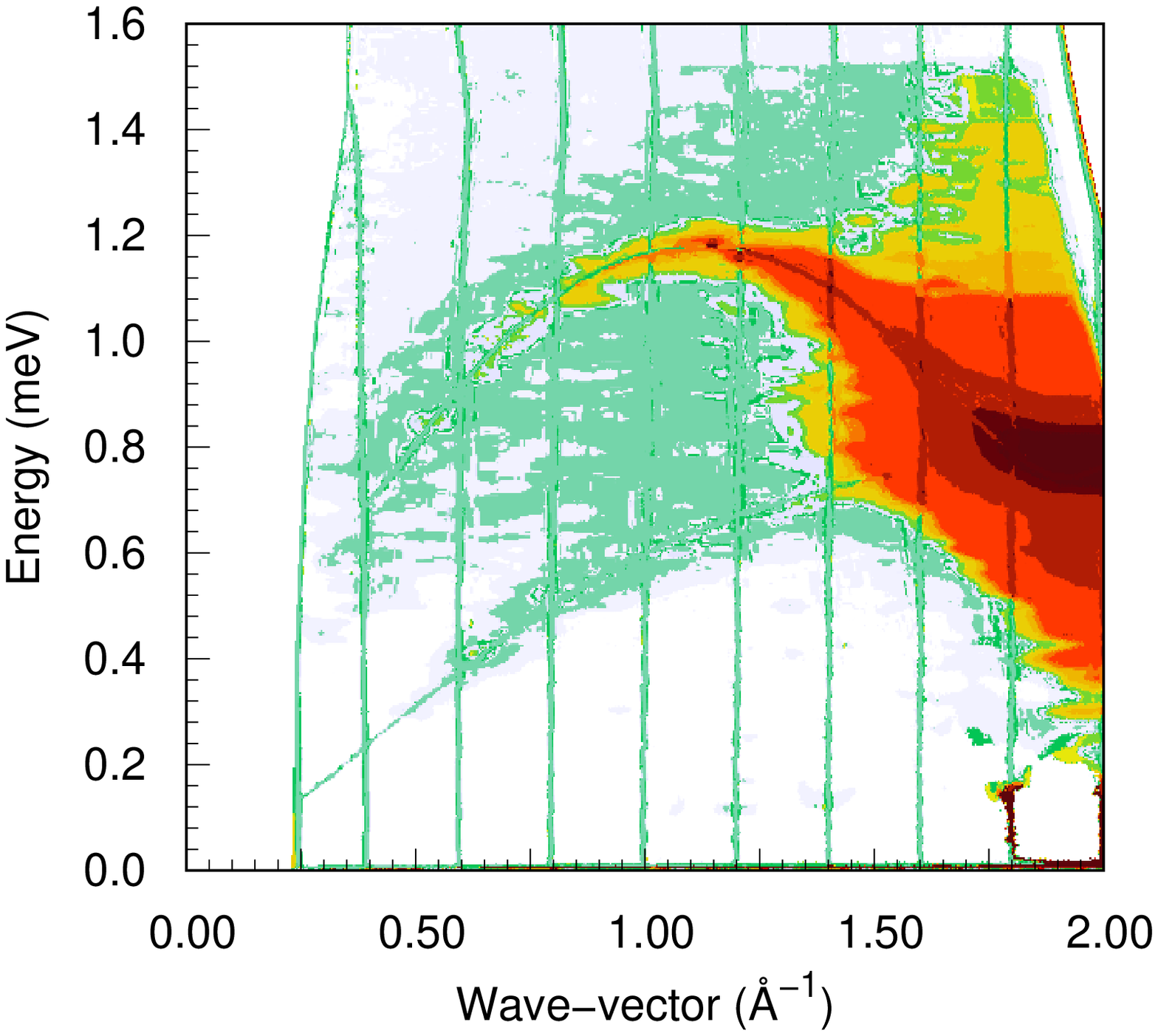}
	\includegraphics[width=0.42\textwidth,angle=-90]{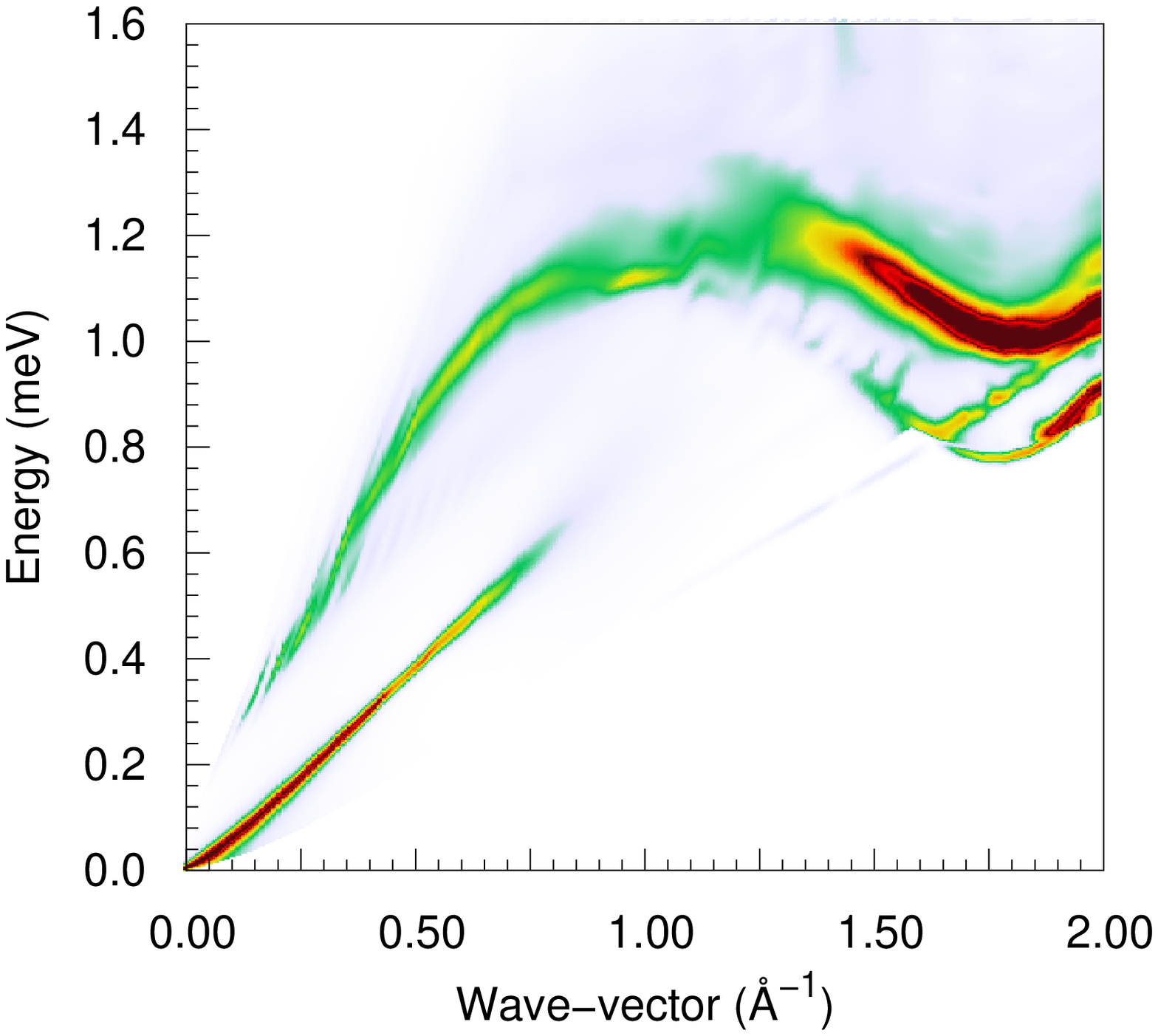}
\caption{\small{Experimental \cite{LauterPRL} (top) and theoretical
    (bottom) \cite{filmexc} determinations of the dynamic structure factor of
    a multilayer film of $^4$He, displaying density and ripplon
    excitations.  \label{fig:2D-4He}}}
 \end{center}
\end{figure}

A typical example is shown in Fig.\,\ref{fig:2D-4He}.  It corresponds
to a multilayer film of about 6 atomic layers, where both density
excitations and ripplons are visible with comparable intensity
\cite{LauterPRL}. The experimental curve is broader than the
theoretical one, an effect largely due to the mosaic spread (angular
distribution) of the graphite substrate, a powder made of microscopic
flat crystals.

Finally, in the sub-monolayer coverage regime, one can observe
gaseous, fluid and solid phases, as well as ``commensurate phases''
where the atoms adopt a periodic arrangement commensurate with the
substrate atomic lattice.

\section{$^3$He in reduced dimensions}

\begin{figure}[H]
   \begin{center}
	\includegraphics[width=0.40\textwidth,angle=-90]{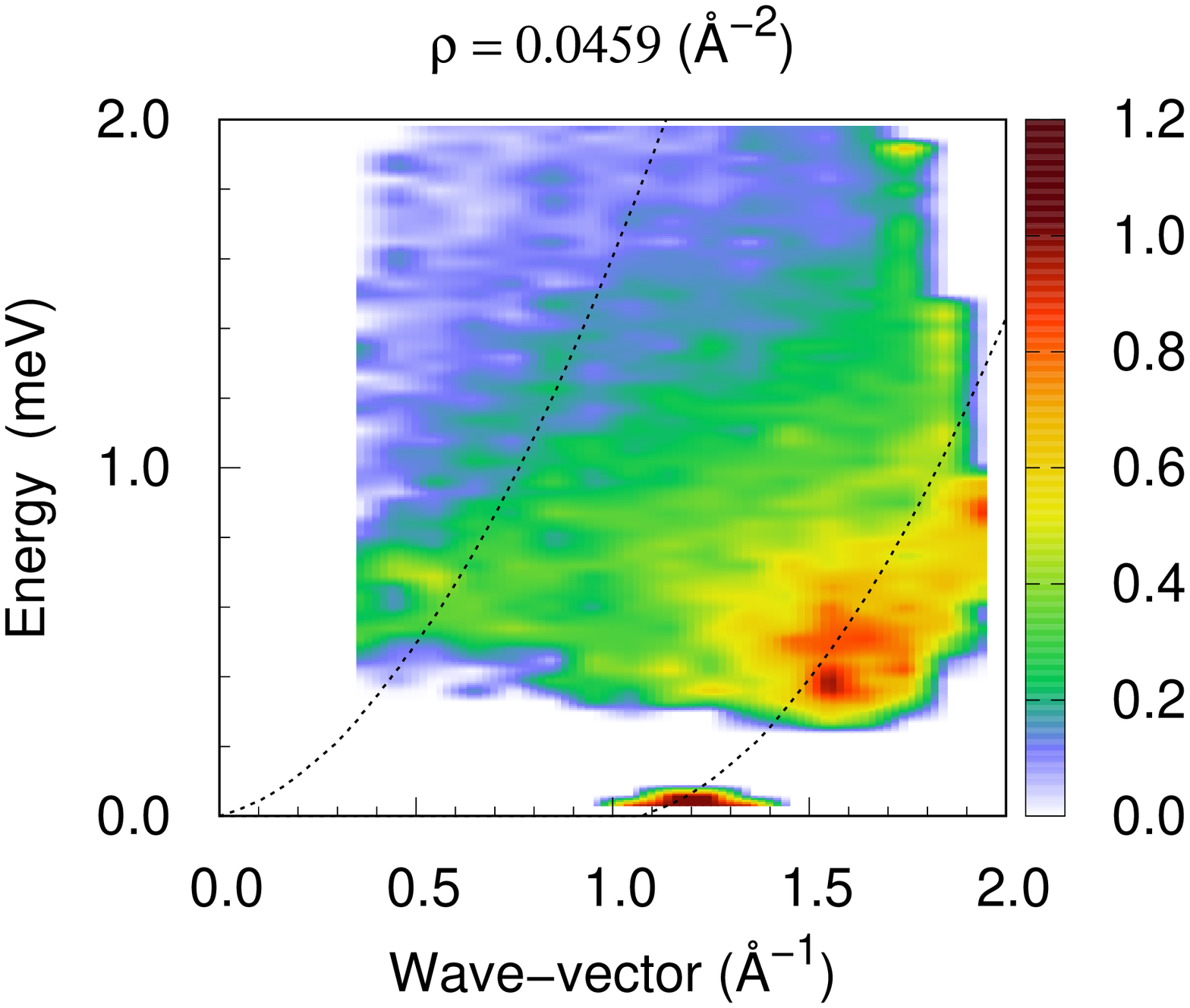}
	\includegraphics[width=0.40\textwidth,angle=-90]{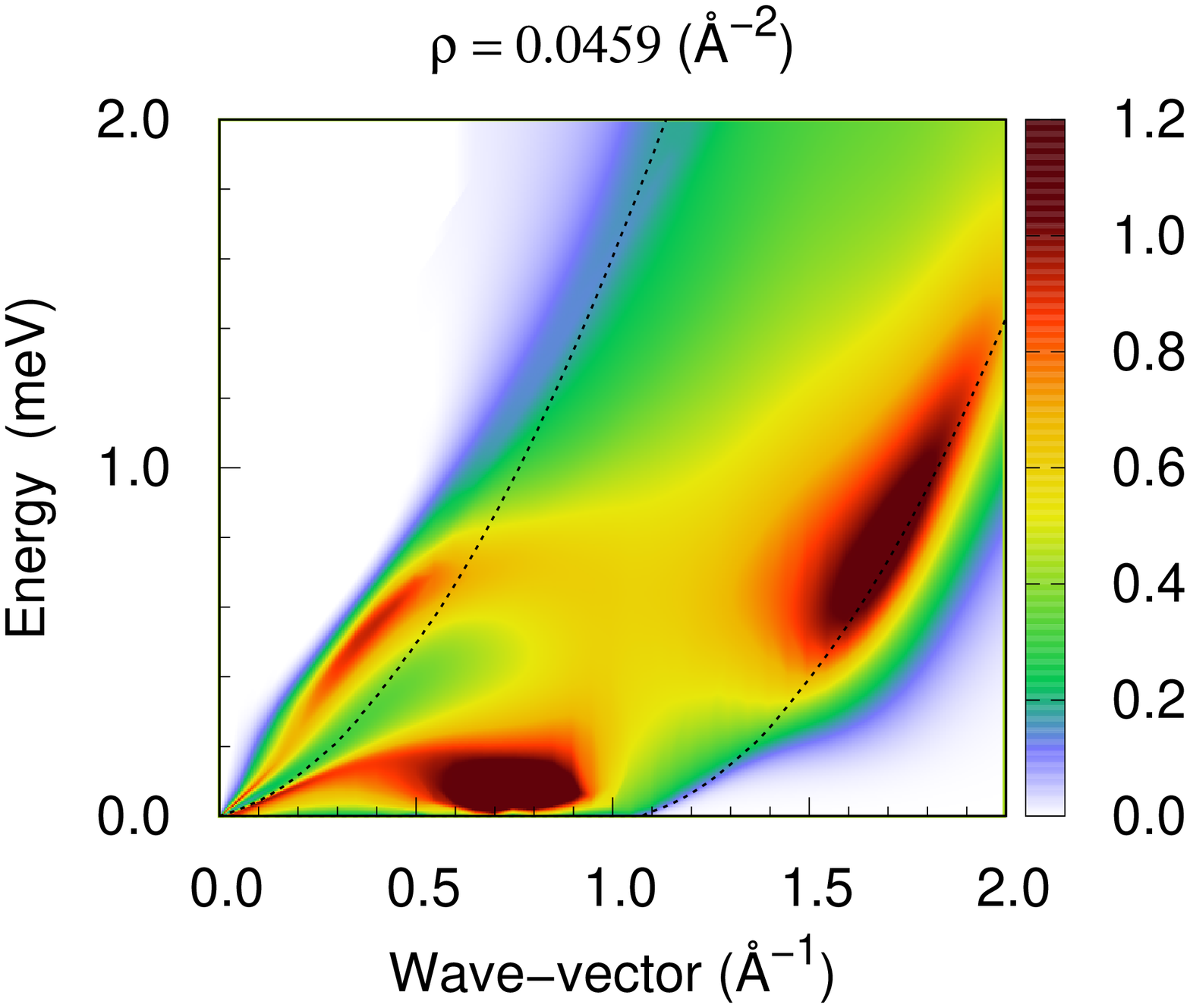}
\caption{\small{Experimental (top) and theoretical (left)
    determinations of the dynamic structure factor of two-dimensional
    liquid $^3$He \cite{Nature_2p2h}. The theoretical calculation
    shows the density and spin-density modes; the latter is not
    visible experimentally, it is masked by a strong elastic signal
    due to the substrate.  A comparison with Fig. \ref{fig:3D-3He}
    shows that the roton-like excitation in the 2D system is found
    outside the particle-hole band (the region limited by the blue
    solid lines), thus avoiding Landau damping. \label{fig:2D-3He}}}
 \end{center}
\end{figure}

The excitations in $^3$He films are shown in
Fig.\,\ref{fig:2D-3He}. They are very similar to those discussed above
for bulk liquid, with the remarkable exception \cite{Nature_2p2h} that
the roton-like excitations are at the edge of the particle-hole band
and hence not strongly affected by Landau damping. These excitations
can therefore propagate, suggesting interesting physical effects
specific to 2D Fermi liquids.

\section{Summary}

The investigation of the dynamics of quantum fluids has been performed
experimentally at very low temperatures using various techniques,
mainly specific heat, compressibility, neutron and X-ray scattering
and, for $^3$He, nuclear magnetic susceptibility. Theoretical
calculations are based on semi-phenomenological approaches
\cite{Aldrich,ALP78,HsuPines}, dynamic many body theory (DMBT)
\cite{2p2h,eomIII,eomIV} as well as different versions of Quantum
Monte Carlo numerical calculations (PIMC, DMC) \cite{BoronatRoton,%
  PhysRevB.93.104510,Arrigoni2013,PhysRevLett.116.135302} to cite only
a few. These have brought a very detailed understanding of these
systems. It is sometimes very difficult to perform experimental
measurements on some particular systems (for instance in reduced
dimensionality), or even impossible to create quite arbitrary systems
("mathematical models" of interaction or external potentials,
etc.). In this case the combination of numerical and microscopic
techniques is essential.

The results find a natural application to many different physical
systems, like nuclear matter, neutron stars, particle physics and
cosmology, where strongly interacting particles are
investigated. Particles themselves, in fact, can be interpreted as
quantized excitations of an underlying field. The helium liquids have
been, in some sense, the Ariadne thread in these investigations: Their
interactions are well established \cite{Aziz,AzizII} and, compared to
nucleon interactions, quite simple. The density of these systems is
very high and they are very ``quantum'', hence they do not permit the
simple treatments that have become popular by the experimental
success to generate and investigate cold quantum gases.  The
experimental challenge is equally high and only the very best of the
experimental equipment and the experimental techniques can stand up to
the challenge. Nevertheless, these challenges have been met on both
the theoretical and experimental side.

\vspace{1cm}

\end{multicols}
\end{document}